\documentclass[11pt,epsf]{article}

\usepackage{epsfig}
\usepackage{amssymb}
\usepackage{graphicx}
\usepackage{color}
\usepackage{subfigure}

\begin{document}

\baselineskip 6mm
\renewcommand{\thefootnote}{\fnsymbol{footnote}}


\newcommand{\nc}{\newcommand}
\newcommand{\rnc}{\renewcommand}



\newcommand{\tcb}{\textcolor{blue}}
\newcommand{\tcr}{\textcolor{red}}
\newcommand{\tcg}{\textcolor{green}}


\def\be{\begin{equation}}
\def\ee{\end{equation}}
\def\ba{\begin{array}}
\def\ea{\end{array}}
\def\bea{\begin{eqnarray}}
\def\eea{\end{eqnarray}}
\def\nn{\nonumber\\}


\def\ct{\cite}
\def\la{\label}
\def\eq#1{(\ref{#1})}


\def\a{\alpha}
\def\b{\beta}
\def\g{\gamma}
\def\G{\Gamma}
\def\d{\delta}
\def\D{\Delta}
\def\ep{\epsilon}
\def\e{\eta}
\def\ph{\phi}
\def\Ph{\Phi}
\def\ps{\psi}
\def\Ps{\Psi}
\def\k{\kappa}
\def\l{\lambda}
\def\L{\Lambda}
\def\m{\mu}
\def\n{\nu}
\def\th{\theta}
\def\Th{\Theta}
\def\r{\rho}
\def\s{\sigma}
\def\S{\Sigma}
\def\ta{\tau}
\def\o{\omega}
\def\O{\Omega}
\def\pr{\prime}


\def\half{\frac{1}{2}}

\def\goto{\rightarrow}

\def\na{\nabla}
\def\grad{\nabla}
\def\curl{\nabla\times}
\def\div{\nabla\cdot}
\def\pa{\partial}

\def\bra{\left\langle}
\def\ket{\right\rangle}
\def\lb{\left[}
\def\lc{\left\{}
\def\ls{\left(}
\def\ln{\left.}
\def\rn{\right.}
\def\rb{\right]}
\def\rc{\right\}}
\def\rs{\right)}

\def\vac#1{\mid #1 \rangle}


\def\td#1{\tilde{#1}}
\def\check{ \maltese {\bf Check!}}


\def\Tr{{\rm Tr}\,}
\def\det{{\rm det}}


\def\bc#1{\nnindent {\bf $\bullet$ #1} \\ }
\def\ch {$<Check!>$ }
\def\ss {\vspace{1.5cm}}

\begin{titlepage}

\hfill\parbox{5cm} { }

\vspace{25mm}

\begin{center}
{\Large \bf  Correlation functions of magnon and spike}

\vskip 1. cm
  { Chanyong Park$^a$\footnote{e-mail : cyong21@sogang.ac.kr}
  and Bum-Hoon Lee$^{ab}$\footnote{e-mail : bhl@sogang.ac.kr}
  }

\vskip 0.5cm

{\it $^a\,$Center for Quantum Spacetime (CQUeST), Sogang University, Seoul 121-742, Korea}\\
{ \it $^b\,$ Department of Physics,, Sogang University, Seoul 121-742, Korea }\\

\end{center}

\thispagestyle{empty}

\vskip2cm


\centerline{\bf ABSTRACT} \vskip 4mm

\vspace{1cm}
After considering a solitonic string moving in the Poincare $AdS$ and two-dimensional sphere,
we calculate the two-point correlation function of magnon and spike
in the semi-classical limit without any explicit solution.
We also calculate the three-point correlation functions
between two heavy and one light operators. We find that the coupling between
two heavy and one light operators  in the string side is the exactly
same as one obtained from the gauge theory by using the RG analysis.

\vspace{2cm}


\end{titlepage}

\renewcommand{\thefootnote}{\arabic{footnote}}
\setcounter{footnote}{0}

\tableofcontents
\section{Introduction}

The conformal field theory (CFT) is characterized by the calculating the conformal dimension
of all primary operators and the structure constant included in the three-point correlation functions,
because higher point functions may be determined by using the operator product expansion (OPE).
The ${\cal N}=4$ super Yang-Mills (SYM) theory in four-dimensional space is an important example to
investigate the interacting CFT \ct{mal1}.  After it was shown that there exists an integrable structure
in ${\cal N}=4$ SYM theory \cite{Beisert:2005tm,Minahan:2002ve,Beisert:2003xu,Beisert:2003yb,Beisert:2004hm,Grigoriev:2007bu,Ahn:2007bq}, there were
great progresses in finding the spectrum of this theory \cite{Ishizeki:2007we}-\cite{Gromov:2009bc}.
This studies were extended to the
ABJM model corresponding to  the low energy theory of M-theory or IIA-string theory
\cite{Aharony:2008ug}-\cite{Abbott:2009um}.
On the contrary, although the structure constant can be evaluated in the weak coupling limit of
SYM by computing the Feynman diagrams,
at the strong coupling there still remain many things to be done.

Recently, there was a big progress in calculating two- and three-point correlation function
semi-classically \cite{Janik:2010gc}.
In the related works \cite{Buchbinder:2010vw}-\cite{Georgiou:2010an}, various two- and three-point
functions for heavy operators
were calculated by using the known explicit solutions.  In this paper, we first revisit the two-point
function of heavy operator like magnon and spike without exact solutions.
Since it is usually very difficult to find the exact string solution corresponding to the heavy operator
in the general set-up, our method would be very helpful to investigate the correlation
function of these heavy operators. Secondly, we also calculate the three-point correlation
function between two heavy and one light operators, where the light operator corresponding to the
marginal scalar operator is dual to a massless scalar field  fluctuation of the dual supergravity theory.
After reading off the structure constant
from the three-point function, we compare it  with the result obtained by the RG analysis in the gauge
theory side.  For magnon described by a spin chain model, we find that
the coupling between two magnons and one marginal scalar operator
at the semi-classical limit of the string sigma model is the exactly same as one obtained from the
RG analysis in the gauge
theory.  Although the dual integrable model describing spike is not clear, if such model exists,
we can identify the spike with another heavy operator in such integrable model, whose conformal
dimension is given by calculating two-point function. Since there is no known explicit solution
for spike, we calculate two- and three-point function by using the equations of motion and boundary
conditions of the spike. In Ref. \cite{Arnaudov:2010kk},  by using the equations of motion of
the solitonic string without imposing the boundary conditions, the two- and three-point functions
for various
strings were calculated.
The method using the equations of motion and boundary conditions would be very powerful to calculate
the various correlation
function in general backgrounds because it is usually very difficult to find the exact solution of
the string worldsheet soliton. Moreover,
by taking the analogy with magnon,  we can easily find the coupling between two spikes and one
marginal scalar operator in the gauge theory side. We find that this result is also the exactly same
as one obtained from the
semi-classical calculation in the string sigma model.

The rest part is organized as follows. After explaining the equations of motion and the boundary
conditions for magnon and spike in Sec. 2, we calculated two- and three-point correlation
functions for magnon in Sec. 3 and for spike in Sec. 4.
We conclude with a brief discussion.

\section{Solitonic string on the Poincare $AdS$ and $S^2$}

Consider a magnon or spike moving in the $AdS_M \times S^2$, which can be usually a
subspace of $AdS_5 \times S^5$. If we consider the global patch
for $AdS$, a string solution corresponding to the magnon or spike is located at the
center of $AdS$. However, in the Euclidean Poincare patch
\be
ds_{AdS}^2 = \frac{1}{z^2}  \ls  dz^2 + d \vec{x}^2 \rs ,
\ee
the string solution can be described by a point-particle moving in $AdS$.
Especially, in the conformal gauge the integration over the string worldsheet is reduced to
the integration over the modular parameter $s$ of the cylinder
\be
\int d^2 \s \to \int_{-s/2}^{s/2}  d \ta \int_{-L}^{L} d\s ,
\ee
where we concentrate on the magnon or spike solution and $\pm L$ imply two ends
of the string worldsheet.  Notice that the magnon and spike solutions described by a
open string corresponds to the half of the closed string.

If we choose the space-like separation in the $AdS$ space,
the string action on $AdS_M \times S^2$ is given by \cite{Lee:2008ui,Lee:2008yq}
\be	\la{act:ph0}
S_{st} = - \frac{T}{2} \int d^2 \s \lb - \frac{(\pa_{\ta} x )^2 + (\pa_{\ta} z )^2 }{z^2}- \ls
\pa_{\ta} \th \rs^2 + \ls  \pa_{\s} \th \rs^2
- \sin^2 \th \lc  \ls  \pa_{\ta} \ph \rs^2 -\ls  \pa_{\s} \ph \rs^2 \rc  \rb ,
\ee
where $\frac{T}{2}$ is a string tension,  $T=\frac{\sqrt{\l}}{2 \pi}$ for $AdS_5 \times S^5$.

The solutions of the $AdS$ part, $z(\ta)$ and $x(\ta)$ are given by
\bea
z(\ta) &=& \frac{R}{\cosh \k \ta} , \nn
x(\ta) &=& R \tanh \k  \ta + x_0 ,
\eea
which is the specific parameterization of a geodesic in $AdS$, $(x(\ta) - x_0)^2 +z(\ta)^2 = R^2$.
From these solutions, the action of the $AdS$ part simplifies to
\be
 \frac{T}{2} \int_{-s/2}^{s/2} d \ta \int_{-L}^{L} d \s \frac{(\pa_{\ta} x )^2 + (\pa_{\ta} z )^2 }{z^2} =
  \frac{T}{2} \int_{-s/2}^{s/2} d \ta \int_{-L}^{L} d \s \k^2 .
\ee
Imposing the boundary conditions
\be
\ls x(-s/2) , z(-s/2) \rs = (0,\ep) \ \  {\rm and } \ \ \ls x(s/2) , z(s/2) \rs = (x_f,\ep) ,
\ee
in which $\ep$ is very small and corresponds to an appropriate UV cut-off,
we can find a relation between $\k$ and $x_f$
\be	\la{rel:kappa}
\k \approx \frac{2}{s} \log \frac{x_f}{\ep} ,
\ee
with $x_f \approx 2 R \approx 2 x_0$.

Now, consider the $S^2$ part of the string solution. Under the following parameterization
\be
\th = \th (y) \ , \ \  \ph = \n \ta + g (y) \ , \ \ {\rm and}  \  y = a \ta + b \s ,
\ee
the equations of motion for $\ph$ reads off
\be
0 = \pa_y \lc \sin^2 \th \ls a \n + (a^2 - b^2) g' \rs \rc ,
\ee
where the prime means the derivative with respect to $y$.
So $g'$ can be rewritten
in terms of $\th$
\be	\la{eq:ph}
g' = \frac{1}{b^2 - a^2} \ls a \n - \frac{c}{\sin^2 \th }\rs ,
\ee
where $c$ is an integration constant. The equation of motion for $\th$ after multiplying $2 \th'$
can be rewritten as the following form
\be
0 = \pa_y \ls \th'^2 + \frac{b^2 \n^2}{(b^2 - a^2)^2} \sin^2 \th + \frac{c^2}{(b^2-a^2)^2 \sin^2 \th} \rs .
\ee
Therefore, we can also rewrite $\th'$ in terms of $\th$
\be 	\la{eq:th}
\th'^2 = \frac{b^2 \n^2}{(b^2 - a^2)^2 \sin^2 \th} \lb - \sin^4 \th +\frac{w^2}{b^2 \n^2} \sin^2 \th
- \frac{c^2}{b^2 \n^2} \rb .
\ee
where $\frac{w^2}{(b^2 - a^2)^2}$ is introduced as another integration constant.

To find exact solutions of
\eq{eq:ph} and \eq{eq:th}  the above two integration parameters $c$ and $w$ should be
fixed by appropriate boundary conditions.
We first impose that $\th$ has a maximum value $\th_{max}$ satisfying $\th_{max}' = 0$, which
plays an important role to determine the size of dual magnon or spike. For example,
if $\sin \th_{max}=1$, the magnon and spike have an infinite energy.
For magnon,  this also implies an infinite angular momentum, which can be reinterpreted
as a infinite size of the spin chain operator in the dual gauge theory.
 Notice that this boundary condition should be imposed to both magnon
and spike, so the above \eq{eq:th} can be rewritten as
\bea
\th'^2 = \frac{b^2 \n^2}{(b^2 - a^2)^2 \sin^2 \th}  \lb \ls \sin^2 \th_{max} - \sin^2 \th \rs
\ls \sin^2 \th - \sin^2 \th_{min} \rs \rb ,
\eea
with
\bea
\sin^2 \th_{max} + \sin^2 \th_{min} &=& \frac{w^2}{b^2 \n^2}  , \nn
\sin^2 \th_{max}  \ \sin^2 \th_{min} &=& \frac{c^2}{b^2 \n^2} .
\eea
Until now, there is no difference between magnon and spike. The difference between them
appears when imposing the  second boundary condition.
For the magnon, we should impose
$\pa_{\s} \ph = 0$ at $\th=\th_{max}$, which guarantees that  even for $\sin \th_{max}=1$
the angle difference is finite while the energy and the angular momentum are infinite. Notice that
this is the typical structure of the magnon's dispersion relation.
By solving this boundary condition, we can find $\sin^2 \th_{max} = \frac{c}{a \n}$. Inserting
this result into \eq{eq:th}, the integration constant $w^2$ becomes
\be	\la{rel:par}
w^2 = \frac{c \n (a^2 +b^2)}{a} .
\ee

Since the spike has usually a finite angular momentum with a infinite energy and angle difference,
we should impose a different
boundary condition from the magnon. For $\sin \th_{max}=1$, to find a finite angular
momentum, we should impose
 $\pa_{\ta} \ph = 0$ at $\th=\th_{max}$.
This boundary condition for spike gives $\sin^2 \th_{max} = \frac{ac}{ \n b^2}$.
Inserting this result into \eq{eq:th}, we can reobtain the above result in \eq{rel:par}. Notice that
\eq{rel:par} is related to one of the Virasoro constraints \cite{Lee:2008ui}, so it should
be satisfied in both cases,  magnon and  spike. As a result, since
 $w^2$ has the same form in the magnon and spike, \eq{eq:th} can be reduced to
\be	\la{eq:thp}
\th'^2 = \frac{b^2 \n^2}{(b^2 - a^2)^2 \sin^2 \th}  \lb \ls \frac{c}{a \n}- \sin^2 \th \rs
\ls \sin^2 \th - \frac{ac}{\n b^2} \rs \rb .
\ee
In the case of $\sin \th_{max} = 1$, the $\sin^2 \th_{min}$ of the magnon solution is given
by $\frac{a^2}{b^2}$. Since $\sin \th_{min}$ is always smaller than $1$, two parameters, $a$ and $b$,
of the magnon should satisfy $b > a$, where we assume that all parameters are positive.
For the spike, since $\sin \th_{max} = 1$ gives $\sin^2 \th_{min}= \frac{b^2}{a^2}$,
the parameter constraint $a > b$ should be satisfied.
Although the solution $\th$
of \eq{eq:thp} has an additional integration constant, it is  irrelevant
to calculate the correlation functions of the magnon and spike.

\section{Magnon}

\subsection{Two-point correlation function }

After convolution with the relevant wave function following the Janik's work \cite{Janik:2010gc}, the new action
is given by
\bea
\bar{S} &=& S - \Pi_{\th} \dot{\th} - \Pi_{\ph} \dot{\ph}  \nn
&=& - \frac{T}{2} \int_{-s/2}^{s/2} d \ta \int_{-L}^{L} d \s \frac{\n c}{a} \equiv - \frac{T}{2}
\int_{-s/2}^{s/2} d \ta \int_{-L}^{L} d \s \ \r^2,
\eea
where $\int d \s \r$ corresponds to the energy of a solitonic string moving on $S^2$ and
$2L$ is the length of the worldsheet string. In the dual gauge theory side,  $\int d \s \r$
and $L$ corresponds to the conformal dimension and the size of the dual spin chain operator,
respectively.
If we take the $L \to \infty$ limit which is the same as taking $\sin \th_{max}=1$,
the worldsheet solitonic string solution corresponds to the spin chain operator having infinite size
in the dual gauge theory. From now on, we concentrate on the infinite size of magnon and spike.
Using the above together with \eq{rel:kappa},
 the total action for the magnon
is given by
\be
i S_{tot} \equiv i \ls S_{AdS} + \bar{S} \rs = i \ls \frac{4}{s^2} \log^2 \frac{x_f}{\ep}  - \r^2 \rs s L T .
\ee
From this total action, the saddle point of the modular parameter $s$ reads
\be	\la{res:spmag}
\bar{s} = - i \frac{2}{\r} \log \frac{x_f}{\ep} ,
\ee
which corresponds to the Virasoro constraint for the einbein.
At this saddle point, $\k$ and $\r$ are related by $\k = i \r$ and the semi-classical partition function
of the giant magnon
becomes
\be	\la{res:scpf}
e^{i S_{tot}}  = \ls  \frac{\ep}{x_f} \rs^{2 E} ,
\ee
where $E$ corresponding to the magnon's conformal dimension is
\be
E = T \int_{-L}^{L}  d \s \r  = 2 T \int_{\th_{min}}^{\pi/2} d \th \frac{\cos^2 \th_{min} \sin \th}{\cos
\th \sqrt{\sin^2 \th - \sin^2 \th_{min}}}.
\ee
Using the definitions for the angular momentum $J$ and the angle difference $\D \ph$, which can be
identified with the string worldsheet momentum $p$, \cite{Lee:2008yq} 
\bea
J &=& T \int_{-L}^{L} d \s \sin^2 \th  \pa_{\ta} \ph
= 2 T   \int_{\th_{min}}^{\pi/2} d \th \frac{\sin \th \ls \sin^2 \th - \sin^2 \th_{min}\rs }{\cos
\th \sqrt{\sin^2 \th - \sin^2 \th_{min}}}  , \nn
| \D \ph | &\equiv& p = - \int d \ph  = 2   \int_{\th_{min}}^{\pi/2} d \th \frac{\sin \th_{min}
\cos \th }{\sin \th \sqrt{\sin^2 \th - \sin^2 \th_{min}}} ,
\eea
the energy $E$ can be rewritten in terms of $J$ and $p$ 
\be
E = J + 2 T  | \sin  \frac{p}{2}|  ,
\ee
which corresponds to the dispersion relation of a magnon in the spin chain model
at the strong t' Hooft coupling
limit. This result shows that the computation of the semi-classical partition function of the
bulk string theory leads to the two-point correlation function
of the dual operator in the gauge theory.

\subsection{Three-point correlation function}

Now, we consider the deformation by a marginal scalar primary operator, whose dual bulk field is
described by a massless scalar field in $AdS_5$. In  Ref. \cite{Costa:2010rz},
it was shown  by using the RG analysis  that the deformed anomalous dimension can be related to
the coupling between
any two operators in the undeformed theory and the  marginal scalar operator in the dual field
theory side. Furthermore,
it was also shown that the same result can be also obtained by calculating the partition function of the
string theory in the semi-classical limit, in which the known exact solutions were used.

In this section, we revisit the three-point correlation function between two heavy magnon operators
and one light operator corresponding to the marginal scalar operator without any explicit solution.
The bulk-to-boundary propagator of a massless scalar field $\chi$ in $AdS$ is given by
\cite{Freedman:1998tz}
\be
K_{\chi} (x^{\m},z;y^{\n}) = \frac{6}{\pi^2 } \ls \frac{z}{z^2 + (x-y)^2 } \rs^4 .
\ee
Then, the three-point function  between two magnon operators denoted by  ${\cal  O}_m$ and marginal
scalar operator ${\cal D}_{\chi}$ is given by \cite{Costa:2010rz}
\be	\la{for:three}
\bra {\cal O}_m (0) {\cal O}_m (x_f)  {\cal D}_{\chi} (y) \ket \approx
\frac{I_{\chi}[\bar{X},\bar{s};y]}{|x_f|^{2 E}}  ,
\ee
with
\be
I_{\chi}[X,s;y] = i \int_{-s/2}^{s/2} d\ta \int_{-L}^{L} d \s \ln \frac{\d S_p[X,s,\chi]}{\d \chi}
\right|_{\chi=0} \ K_{\chi} \ls X(\ta,\s);y \rs ,
\ee
where $\chi$ corresponds to the massless dilaton fluctuation and $S_p[X,s,\chi]$ represents
the Polyakov action including the dilaton fluctuation
\be
S_p[X,s,\chi] = - \frac{T}{2} \int d^2 \s \ \sqrt{- \g} \g^{\a\b} \pa_{\a} X^A 
\pa_{\b} X^B G_{AB} \ e^{\chi/2} + \cdots .
\ee
For the magnon case, $I_{\chi}[X, s ;y]$ becomes
\bea
I_{\chi}[X,s;y] = i \frac{3}{\pi^2} S_{st} \times  \ls \frac{z}{z^2 + (x-y)^2 } \rs^4 ,
\eea
where $S_{st}$  in  \eq{act:ph0} is the Polyakov action in the absence of the dilaton field.
Inserting solutions obtained in
the previous section, the above integration is reduced to
\bea \la{int:I}
I_{\chi}[X,s;y] &=& i \frac{3 T}{2 \pi^2} \int_{-s/2}^{s/2} d\ta \int_{-L}^{L} d \s
\lb \k^2 + \frac{1}{b^2 -a^2} \ls 2 b^2 \n^2 \sin^2 \th - \frac{\n c}{a} (a^2 + b^2) \rs \rb \nn
&& \qquad \qquad \qquad \qquad \times  \ls \frac{z}{z^2 + (x-y)^2 } \rs^4 .
\eea
Notice that $\k = i \r$ at the saddle point and that the propagator of the massless field  depends
on $\ta$ only.  For the infinite size magnon case $\sin^2 \th_{max} =1$, the integration over $\s$,
after using the chain rule twice,
becomes
\bea
 \int_{-L}^{L} d \s \lb \k^2 + \frac{1}{b^2 -a^2} \ls 2 b^2 \n^2 \sin^2 \th
- \frac{\n c}{a} (a^2 + b^2) \rs \rb
&=& - 4 \r  \int_{\th_{min}}^{\th_{max}} d \th \frac{\sin \th \cos \th}{\sqrt{\sin^2 \th
- \sin^2 \th_{min}}} \nn
&=& - 4 \r \cos \th_{min}
\eea
where we use $c=a \n$ and $\r=\n$ which are satisfied  only for $\sin^2 \th_{max} =1$.
The integration $I_{\chi}[X,s;y]$ at the saddle point is reduced to
\bea
I_{\chi}[\bar{X},\bar{s};y] &=& - i \frac{6 T }{\pi^2} \r \cos \th_{min}  \int_{-s/2}^{s/2} d \ta
 \ls \frac{z}{z^2 + (x-y)^2 } \rs^4 \nn
 &\approx& - \frac{T }{2 \pi^2}  | \sin \frac{p}{2} |  \frac{x_f^4}{y^4 (x_f-y)^4} .
\eea
Therefore, the three-point correlation function from \eq{for:three} becomes
\be
\bra {\cal O}_m (0) {\cal O}_m (x_f)  {\cal D}_{\chi} (y) \ket = - \frac{T }{2 \pi^2}  | \sin \frac{p}{2} |  \frac{1}{x_f^{2 E - 4} y^4 (x_f-y)^4} .
\ee
From this result, the coupling $a_{{\cal D}AA}$ reads off
\be	\la{res:coupling}
2 \pi^2 a_{{\cal D}mm} = - T |\sin \frac{p}{2}|
\ee
In the integrable spin chain model, the conformal dimension of the magnon is given
by
\be
\D = J + \sqrt{1 + 16 g^2  | \sin \frac{p}{2} |^2 } ,
\ee
where the  string tension $T$ can be rewritten in terms of the 't Hooft coupling $\l$
\be
T = 2 g  \ \ {\rm and} \ \ g^2 = \frac{y_{YM}^2 N}{16 \pi^2} = \frac{\l}{16 \pi^2} .
\ee
From this magnon conformal dimension, we can extract the coupling between two magnon operators
and one marginal scalar operator  by using
the following formula obtained by the RG analysis \cite{Costa:2010rz}
\be	\la{re:thmag}
2 \pi^2 a_{{\cal D}mm}  = - g^2 \frac{\pa}{\pa g^2} \D = - \frac{8 g^2  | \sin
\frac{p}{2} |^2}{\sqrt{1 + 16 g^2  | \sin \frac{p}{2} |^2 }} .
\ee
In the large 't Hooft coupling limit, the result \eq{re:thmag} in the gauge theory is reduced to
one of the string calculation \eq{res:coupling} in the semi-classical limit.

\section{Spike}

\subsection{Two-point correlation function}

Now, we perform the similar calculation for the spike. Since there is no known explicit
solution, the method explained in the previous sections would be very useful.
For the spike, since $a>b$, it is more convenient to write $\th'^2$ as the following form
\be	\la{eq:thsp}
\th'^2 = \frac{b^2 \n^2}{(a^2 - b^2)^2 \sin^2 \th}  \lb \ls \sin^2 \th_{max}- \sin^2 \th \rs
\ls \sin^2 \th - \sin^2 \th_{min} \rs \rb .
\ee\
with
\be
\sin^2 \th_{max} = \frac{ac}{b^2 \n}  \ \ {\rm and} \ \ \sin^2 \th_{min} = \frac{c}{a \n} .
\ee
Here, we concentrate on the infinite size case, $\sin^2 \th_{max} =1$. In this case,
since $c/\n=b^2/a$, $\sin^2 \th_{min} = \frac{b^2}{a^2}$. Like the magnon case, the
new convoluted action $\bar{S}$ is also given by
\be	\la{rel:rho}
\bar{S} = - \frac{T}{2} \int d^2 \s \r^2 = - \frac{T}{2} \int d^2 \s \frac{\n c}{a} .
\ee
Using this result, we can find the same saddle point  \eq{res:spmag} and the semi-classical
partition function as ones obtained in the magnon  case
\be
e^{i S_{tot}}  = \ls  \frac{\ep}{x_f} \rs^{2 E} ,
\ee
with
\be
E = T \int_{-L}^{L}  d \s \r  = 2 T \int_{\th_{min}}^{\pi/2} d \th \frac{\cos^2 \th_{min} \sin \th}{
\sin \th_{min} \cos \th \sqrt{\sin^2 \th - \sin^2 \th_{min}}}.
\ee
Using the conserved charges and angle difference \cite{Lee:2008yq}
\bea
J &=& T \int_{-L}^{L} d \s \sin^2 \th  \pa_{\ta} \ph
= 2 T   \int_{\th_{min}}^{\pi/2} d \th \frac{\sin \th \cos \th }{\sqrt{\sin^2 \th
- \sin^2 \th_{min}}}  , \nn
| \D \ph | &\equiv& p = - \int d \ph  = 2   \int_{\th_{min}}^{\pi/2} d \th \frac{\sin^2 \th
-\sin^2 \th_{min} }{\sin \th_{min} \sin \th \cos \th \sqrt{\sin^2 \th - \sin^2 \th_{min}}} ,
\eea
the dispersion relation of the spike
can be written as 
\be	\la{rel:sp1}
E = T \D \ph + 2 T \td{\th},
\ee
with
\be 	\la{rel:sp2}
\td{\th} \equiv \frac{\pi}{2} - \th_{min} = \arcsin \frac{J}{2 T} .
\ee

\subsection{Three-point correlation function}

Now, we consider the three-point correlation function between two spike operators and one
light operator dual to a
massless scalar field $\chi$. From the formula for the three-point function in \eq{for:three},
we can find
\bea
I_{\chi}[X,s;y] &=& i \frac{3}{\pi^2} S_{st} \times  \ls \frac{z}{z^2 + (x-y)^2 } \rs^4 ,
\eea
where $S_{st}$ is also given by \eq{act:ph0}. Inserting the spike solution into this string action,
at the saddle point \eq{res:spmag} where $\k=i\r$ is satisfied, the string action is reduced to
\be
S_{st} = \frac{T}{2} \int_{-\bar{s}/2}^{\bar{s}/2} d \ta \int_{-L}^{L} d \s \frac{2 b^2 \n^2}{a^2 - b^2}
\ls \frac{c}{a \n} - \sin^2 \th \rs.
\ee
In terms of the $\th$ integration instead of $\s$, the string action can be rewritten as
\bea
S_{st} &=&  - T  \int_{-\bar{s}/2}^{\bar{s}/2} d \ta \frac{b \n}{a} \D \ph
 +  \n T \int_{-\bar{s}/2}^{\bar{s}/2} d \ta \int_{\sin^2 \th_{min}}^{\sin^2 \th_{max}=1} d x
 \frac{x-\sin^2 \th_{min}}{x \sqrt{x-\sin^2 \th_{min}}} \nn
 &=&   \n T \sin \th_{min} \int_{-\bar{s}/2}^{\bar{s}/2} d \ta  \lb  - \D \ph + 2  \lc \cot \th_{min}
 - \arctan(\cot \th_{min} ) \rc \rb.
\eea
Then, $I_{\chi}[\bar{X},\bar{s};y]$ at the saddle point $s=\bar{s}$ becomes
\be
I_{\chi}[\bar{X},\bar{s};y] = \frac{T}{4 \pi^2}
\lb -  \D \ph + 2  \lc \cot \th_{min} - \arctan(\cot \th_{min} ) \rc \rb \frac{x_f^4}{y^4 (x_f- y)^4} ,
\ee
where we use $\r = \sqrt{\frac{\n c}{a}}$ in \eq{rel:rho}. In terms of $\td{\th}$,
$I_{\chi}[\bar{X},\bar{s};y]$
can be rewritten as
\be
I_{\chi}[\bar{X},\bar{s};y]  = - \frac{T}{4 \pi^2}
\lb   \D \ph - 2  \lc \tan \td{\th} - \td{\th}  \rc \rb \frac{x_f^4}{y^4  (x_f- y)^4} .
\ee
Finally, the three-point correlation function becomes
\be
\bra {\cal O}_s (0) {\cal O}_s (x_f)  {\cal D}_{\chi} (y) \ket = \frac{1}{2 \pi^2} 
\ls  - \frac{T \D \ph }{2}
 -  T \td{\th}   + T  \tan \td{\th}  \rs    \frac{1}{x_f^{2 E - 4} y^4  (x_f- y)^4}  ,
\ee
where ${\cal O}_s$ implies the dual spike operator.
From this three-point function, we easily read off the coupling
\be	\la{res:csp}
2 \pi^2 a_{{\cal D}ss} = - \frac{T \D \ph }{2}
 -  T \td{\th}   + T  \tan \td{\th} .
\ee
Although the dual integrable model of the spike solution is not clear, it is widely believed
that there exists a dual integrable model related to spike.  Keeping this in mind, we can say that
the dual operator of the spike has the dispersion relation in \eq{rel:sp1} with the relation
\eq{rel:sp2}.
By taking the analogy with the magnon, we can calculate the coupling
between two spikes and one marginal scalar operator in the gauge theory side
\bea
2 \pi^2 a_{{\cal D}ss}  &=& - g^2 \frac{\pa}{\pa g^2} E = - \frac{T}{2} \frac{\pa}{\pa T} \ls T \D \ph
+ 2 T \arcsin \frac{J}{2 T} \rs \nn
&=&  - \frac{T \D \ph }{2}  -  T \td{\th}   + T  \tan \td{\th}  ,
\eea
which is the exactly same as  \eq{res:csp} calculated by the string partition function at the semi-classical limit.

\section{Discussion}

We calculated the two- and three-point correlation functions of magnon and spike without
any explicit solution. Here, instead of finding the explicit solutions, we used the equations
of motion and boundary conditions. After calculating the three-point correlation function
between two heavy magnons and one light operator, which is dual to one of the massless scalar field
of various supergravity fields, we extracted the structure constant describing the coupling between
those operators. This structure constant
calculated in the string set-up is the exactly same as ones obtained by the RG analysis in the gauge
theory side .
We also calculated the two- and three-point correlation function
between two spikes and one light operator. Although the dual integrable model for spike
is not clear, by taking the analogy with the magnon we have calculated the structure constant in the
gauge theory side. We also saw that this result is consistent with one
calculated from the string sigma model.

In more general backgrounds, since it is usually very
difficult to find the exact solution, our method would be very helpful to investigate various
correlation functions for the solitonic string solutions like magnon and spike .

\vspace{1cm}

{\bf Acknowledgement}

CP thanks to C. Ahn for helpful discussion.
This work was supported by the National Research Foundation of Korea(NRF) grant funded by
the Korea government(MEST) through the Center for Quantum Spacetime(CQUeST) of Sogang
University with grant number 2005-0049409. C. Park was also
supported by Basic Science Research Program through the
National Research Foundation of Korea(NRF) funded by the Ministry of
Education, Science and Technology(2010-0022369).

\end{document}